\documentclass[12pt]{amsart}

\usepackage{epsfig}

\begin{document}

\title[]{\bf Conformal Scalar Propagation and Hawking Radiation}

\author[]{George Tsoupros \\
       {\em Private address - Beijing}\\
       {\em People's republic of China}\\
}
\thanks{present e-mail address: landaughost@hotmail.com}

\begin{abstract}

The construction of the conformal scalar propagator which has been
obtained in the preceding two projects as an analytic function of
the Schwarzschild black-hole space-time is completed with a boundary
condition imposed by the physical context through contour
integration in the exterior vicinity of the event horizon. It is
shown that, as a consequence of the semi-classical character which
the emitted quanta have in that exterior vicinity, the particle
production by the Schwarzschild black hole which was formally
established in the preceding project is identical to thermal Hawking
radiation. By extension, it is established that such a particle
production corresponds to a spectrum which detracts from thermality
by the amount predicted by Parikh and Wilczek if energy conservation
is properly imposed as a constraint on scalar propagation. The
results obtained herein support the case made by S. Hawking on the
relation between quantum propagation and observation of particles
produced by a black hole.

\end{abstract}

\maketitle

``Du siehst, mein Sohn, zum Raum wird hier die Zeit!"

(``You see my son, in here time becomes space!")

\hspace{0.9in} From Richard Wagner's opera ``Parsifal"


{\bf I. Introduction}\\

The transition amplitude between two events is essential to the
investigation of a quantum field's dynamical behaviour in any curved
space-time. The primary reason is the intimate relation which exists
on general theoretical grounds between the Feynman propagator and
the expectation value $ <T_{\mu\nu}>$ in any specific vacuum state.
In a black-hole space-time the Feynman propagator receives
additional significance if only because it provides a concrete
mathematical description of the consolidated intuitive approach to
the original rigorous frequency-mixing derivation of black-hole
radiation \cite{Stephen}. That approach advances on the heuristic
description of vacuum activity in terms of the transient and
unobservable existence of positive-energy pairs of a particle and
its anti-particle, very much in the spirit of Dirac's old hole
theory. Since, with respect to spatial infinity, the interior region
of the black hole necessarily contains negative-energy states a
positive-energy particle changes to a negative-energy one upon
crossing the event horizon. In effect, the ``uncertainty principle"
inherent in this heuristic approach does not necessitate
re-annihilation in the event that one of the two particles produced
by vacuum fluctuations in the exterior region crosses the event
horizon. The ``uncertainty principle" is still identically upheld
if the remaining positive-energy particle transits to spatial
infinity or else registers its existence on a detector situated at
any distance above the event horizon. For a Schwarzschild black
hole of mass $ M$ in Plank units $ (G = c = \hbar = k = 1)$ this
heuristic approach to Hawking radiation yields the same rigorously
established thermal spectrum which corresponds to the Hawking
temperature

$$
T_H = \frac{1}{8\pi M}
$$

Inherent in this result is the assumption that the background
geometry for $ r \geq 2M$ remains fixed. Such a semi-classical
approximation is, in turn, justified by the fact that the rate of
change of the metric $ \sim M^{-3}$ is very slow compared to the
typical frequency of the radiation $ \sim M^{-1}$ as long as $ M$ is
much bigger than the Planck mass. It would, for that matter, be a
valid approximation to neglect the time dependence of the metric and
to calculate the emission from a sequence of stationary metrics
\cite{Hawking}.

The thermal spectrum which emerges as a result of the stated
semi-classical approximation lies at the core of the centrally
important issue of unitary evolution in the presence of event
horizons. An alternative approach to black-hole radiation which is
based entirely on the above-stated heuristic pair production as an
expression of vacuum fluctuations in a Schwarzschild black-hole
space-time yields a correction to the thermal spectrum by dismissing
the comparison between the metric's rate of change and the frequency
of the emitted radiation in favour of energy conservation for each
emitted quantum particle \cite{Wilczek}. That approach has the
distinctive feature of treating each emitted quantum particle as the
result of such tunneling across the event horizon as is
characterised by a barrier which crucially depends on the tunneling
particle itself \cite{Parikh}. The detraction from thermality
emerges as a consequence of the fact that energy conservation causes
the event horizon to contract to a smaller Schwarzschild radius in
the event of emission. In effect, in the exterior vicinity of the
event horizon the geometry is invariably dynamic. An immediate
consequence of such an approach is the fact that the emitted
particles are also the result of tunneling from the interior region
across the event horizon forward in each particle's proper time, a
feature which is absent in the heuristic treatment of black-hole
radiation based on the stated semi-classical approximation.

Certain features of the heuristic pair-production approach to
Hawking radiation - both, in the context of the assumption of a
semi-classical geometry and in that of a dynamic geometry - deserve
attention in their own merit. First - contrary to a common claim -
the radiation emitted by a black hole is not a consequence of the
gravitational field. It is, instead, the exclusive consequence of
space-time's causal structure determined by the presence of the
event horizon. This general property of black-hole radiation is
inherent in the preceding analysis and is more rigorously analyzed
in \cite{Visser}. Second, such an approach is relevant only to the
Hartle-Hawking and to the Unruh vacuum states respectively. Its
character as heuristic and intuitive stems from the fact that it
does not define those vacuum states but is, instead, the result of
those states' instability. Third, a crucial aspect of the
pair-production approach is the fact that - as a consequence of
infinite time dilation - observers situated close to the event
horizon in decreasing order of the Schwarzschild radial coordinate
necessarily register each particle emitted by the black hole in an
increasingly semi-classical state of very high energy. In the
vicinity of the event horizon itself the corresponding uncertainties
in the probability distributions of each emitted particle tend to
zero. Consequently, in the vicinity of the event horizon the
semi-classical limit $ \hbar \rightarrow 0$ naturally characterises
the emitted particles. This is an equivalent semi-classical
description to that which underlies the derivation in
\cite{Wilczek}.

It is precisely this third feature which shall constitute the basis
of the following analysis. In \cite{George} and \cite{Tsoupros} the
propagator for a massless conformal scalar field in a Schwarzschild
black-hole space-time has been obtained in the Hartle-Hawking state
for a finite range of values of the Schwarzschild radial coordinate
above and below the Schwarzschild radius $ r_S = 2M$ respectively.
That propagator has the unique feature of being an explicitly
analytic function of the background space-time geometry. As that
propagator coincides with the exact propagator on the entire
Schwarzschild black-hole geometry only in that finite range of
values it is ideally suited for the description of observations made
in a local frame established close to the event horizon. In fact, by
properly matching the massless conformal scalar propagator in the
interior region of the Schwarzschild black hole with that in the
exterior region across the event horizon it was formally established
in the heuristic context of pair production that, within a finite
advance of his proper time, an observer situated at any $ r
> 2M$ registers particles emitted by the Schwarzschild black
hole \cite{Tsoupros}. In what follows such a physical effect will be
rigorously established. It will be shown that - always with respect
to such an observer - the propagator established in \cite{George}
and \cite{Tsoupros} results in both, the thermal spectrum of Hawking
radiation if for $ r \geq 2M$ the geometry is treated as static in
the context of the stated semi-classical approximation and the
corrected spectrum obtained in \cite{Wilczek} if energy conservation
is accordingly enforced. For that matter, these results will be
identically valid also for the ``inertial" observer at spatial
infinity $ r \rightarrow \infty$.

A classic calculation which derives black-hole radiation and its
thermal spectrum in terms of a transition amplitude in a
Schwarzschild black-hole space-time has been accomplished
in \cite{Stephen Hawking}. However, the inherent
semi-classicality which characterises the emitted
particles close to the event horizon is not manifest. As the
analysis herein will advance on the necessary match of the
expressions in \cite{Tsoupros} and \cite{George} across the
horizon that inherent semi-classicality will be explicit. The
semi-classical character of the emitted particles will, in
turn, constitute the basis for the derivation of both, the thermal
spectrum of black-hole radiation in the absence of horizon-related
evolution and in the detraction from the thermal spectrum which
energy conservation enforces.

An aspect of central importance in the ensuing analysis is the
boundary condition which stems from the physical demand that the
emitted particles correspond to positive-energy states propagating
forward in time. It will be shown that such a boundary condition
constitutes an essential aspect of the massless conformal scalar
propagator and that it completes that propagator's construction
which was achieved in \cite{George} and \cite{Tsoupros} on the
Schwarzschild black-hole geometry.

The analytic expressions obtained for the propagator in
\cite{George} and, especially, in \cite{Tsoupros} are particularly
involved and apparently intractable. However, in addition to the
calculation which established black-hole radiation in
\cite{Tsoupros} the present derivation will also provide yet another
example as to how, despite appearances, the causal structure of the
Schwarzschild black-hole space-time can substantially simplify the
relevant calculations.

{\bf II. The Thermal Spectrum}

The transition amplitude for a field $ \Phi$ between two
events in any space-time is equivalent to the functional integral over
all paths which intercept the associated two space-time points. In Euclidean
signatures obtained by analytically extending the temporal coordinate to imaginary
values this is

\begin{equation}
W_E = \int D[\Phi]e^{-S[\Phi, \nabla_{\alpha}\Phi]}
\end{equation}
where in the positive definite action functional $ S[\Phi, \nabla_{\alpha}\Phi]$
the gradient $ \nabla_{\alpha}$ is defined with respect to the space-time metric
$ g_{\mu\nu}$. At the semi-classical limit $ \hbar \rightarrow 0$ (1) reduces to
the expansion

\begin{equation}
W_E = e^{-S_0[\Phi_0, \nabla_{\alpha}\Phi_0] + O(\hbar)}
\end{equation}
with $ S_0[\Phi_0, \nabla_{\alpha}\Phi_0]$ being the classical action and with
$ O(\hbar)$ describing higher-loop corrections to it. The physical content
inherent in (1) and (2) will constitute the basis of the following
analysis.

The Schwarzschild metric is

\begin{equation}
ds^2 = -(1 - \frac{2M}{r})dt^2 + (1 - \frac{2M}{r})^{-1}dr^2 +
r^2(d\theta^2 + sin^2\theta d\phi^2)
\end{equation}
The analytical extension $ \tau = + it$ of the real-time coordinate
$ t$ in imaginary values results in a Euclidean, positive definite
metric for $ r > 2M$. The apparent singularity which persists at $ r
= 2M$ can be removed by introducing the new radial coordinate

\begin{equation}
\rho = 4M\sqrt{1 - \frac{2M}{r}}
\end{equation}

Upon replacing

\begin{equation}
\beta = 4M
\end{equation}
the metric in the new coordinates is

\begin{equation}
ds^2 = \rho^2(\frac{1}{\beta^2})d\tau^2 +
(\frac{4r^2}{\beta^2})^2d\rho^2 + r^2(d\theta^2 + sin^2\theta
d\phi^2)
\end{equation}

Through the identification of $ \frac{\tau}{4M}$ with an angular
coordinate of period $ 2\pi$ the coordinate singularity at $ r=2M$
corresponds to the origin $ \rho = 0$ of polar coordinates and is
thus trivially removed. This procedure results in a complete
singularity-free positive definite Euclidean metric which is
periodic in the imaginary-time coordinate $ \tau$. The period $ 8\pi
M$ of that coordinate is the underlying cause of the thermal
radiation at temperature $ T = (8\pi M)^{-1}$ emitted by the
Schwarzschild black hole.

By imposing the requirement $ \rho^2 << \beta^2$ the propagator for
a massless conformal scalar field $ \Phi$ in the Hartle-Hawking state
expressed as an explicitly analytic function of the exterior region
of the Schwarzschild black-hole space-time for a certain range of
values of the radial variable $ r$ above $ r = 2M$ has been established
in \cite{George} to be

$$
D(x_2 - x_1) =
$$

$$
\frac{2}{\beta}\frac{1}{\sqrt{\rho_1\rho_2}}\sum_{l=0}^{\infty}\sum_{m=-l}^{l}Y_{lm}(\theta_2,
\phi_2)Y_{lm}^*(\theta_1, \phi_1)\sum_{p =
0}^{\infty}e^{i\frac{p}{\beta}(\tau_2 -
\tau_1)}\int_{u_0[p]}^{\infty}du\frac{cos[\frac{\pi}{4\beta}(4u + 2p
+ 3)(\rho_2 - \rho_1)]}{\pi^2u^2 + 4(l^2 + l + 1)}
$$

$$
- \frac{2}{\beta^{\frac{3}{2}}}\frac{1}{\sqrt{\rho_1}}\times
$$

$$
\sum_{l = 0} ^{\infty}\sum_{m = -l}^{l}\sum_{p =
0}^{\infty}\int_{u_0'[p]}^{\infty}du\frac{cos[\frac{\pi}{4\beta}(4u
+ 2p + 3)(\beta - \rho_1)]}{\pi^2u^2 + 4(l^2 + l +
1)}\frac{J_p(\frac{2i}{\beta}\sqrt{l^2 + l +
1}\rho_2)}{J_p(2i\sqrt{l^2 + l + 1})}e^{i\frac{p}{\beta}(\tau_2 -
\tau_1)}Y_{lm}(\theta_2, \phi_2)Y_{lm}^*(\theta_1, \phi_1) ~~~;~~
$$

\begin{equation}
u_0 >> p ~~~~;~~~ u_0' >> p ~~~;~~~ \frac{\pi u}{\beta}\rho_{2,1}
>> p
\end{equation}
with a range of validity specified by

\begin{equation}
0 \leq \rho_i^2 \leq \frac{\beta^2}{100} ~~~~ ; ~~~~ i = 1, 2
\end{equation}
or, equivalently, by

\begin{equation}
2M \leq r \leq 2.050M
\end{equation}

The expression which amounts to the first term on the right side in (7) is the
singular part $ D_{as}(x_2 - x_1)$ of $ D(x_2 - x_1)$. In terms of the
eigenvalues and eigenfunctions which the associated elliptic operator has in
the Euclidean sector of the Schwarzschild metric $ D_{as}(x_2 - x_1)$
constitutes the asymptotic expression of $ D(x_2 - x_1)$ and as such contains
the singularity at the coincidence space-time limit $ x_2 \rightarrow x_1$
\cite{George}. The expression which amounts to the second term constitutes
the boundary part $ D_b(x_2 - x_1)$ of $ D(x_2 - x_1)$ which ensures the
boundary condition of vanishing propagation on the boundary hypersurface
$ \rho = \beta$ of the Euclidean Schwarzschild black-hole geometry.

In the interior region $ r < 2M$ of (3) the analytical extension to imaginary time
also involves the extension $ \theta \rightarrow i\tilde{\theta}$ as a result of
which (3) reduces to the negative-definite \cite{Tsoupros}

\begin{equation}
ds^2 = -\big{[}[\frac{2M}{r} -  1]d\tau^2 + \frac{1}{\frac{2M}{r} -
1}dr^2 + r^2(d\tilde{\theta}^2 + sinh^2\tilde{\theta}d\phi^2)\big{]}
\end{equation}

Through the analytical extension of (7) into the interior region of
the Schwarzschild black hole the massless conformal scalar
propagator $ D^{(int)}(x_2 - x_1)$ has been established in
\cite{Tsoupros} for a certain range of values of $ r$ below $ r =
2M$. In view of the extensive character of that Green function and
of the fact that only its singular part is relevant to the ensuing
analysis the associated boundary part $ D_b^{(int)}(x_2 - x_1)$
shall not be cited. With $ \tau = \pm i\xi$ and $ \rho = \pm i\zeta$
that singular part itself is

$$
D_{as}^{(int)}(x_2 - x_1) =
\frac{i}{\beta}\frac{e^{-\frac{3\pi}{4\beta}(\zeta_2 -
\zeta_1)}}{\sqrt{\zeta_1\zeta_2}}\int_{\epsilon \rightarrow
0}^{\infty}d|\tilde{p}|\frac{1}{1 - e^{-2\pi|\tilde{p}|}}
e^{i\frac{|\tilde{p}|}{\beta}(\xi_2 -
\xi_1)}e^{i\frac{\pi}{\beta}\big{[}\frac{|\tilde{p}|}{2} +
u_0[|\tilde{p}|]\big{]}(\zeta_2 - \zeta_1)}\times
$$

$$
\sum_{l = 0}^{\infty}\sum_{m = -l}^{l}Y_{lm}(i\tilde{\theta}_2,
\phi_2)Y^*_{lm}(i\tilde{\theta}_1, \phi_1)
\int_{0}^{\infty}dw\frac{e^{-i\frac{\pi}{\beta}w(\zeta_2 -
\zeta_1)}}{\pi^2(u_0[|\tilde{p}|] - w)^2 - 4(l^2 + l + 1)}
$$

$$
+\frac{i}{\beta}\frac{e^{\frac{3\pi}{4\beta}(\zeta_2 -
\zeta_1)}}{\sqrt{\zeta_1\zeta_2}}\int_{\epsilon \rightarrow
0}^{\infty}d|\tilde{p}|\frac{1}{1 - e^{-2\pi|\tilde{p}|}}
e^{i\frac{|\tilde{p}|}{\beta}(\xi_2 -
\xi_1)}e^{i\frac{\pi}{\beta}\big{[}\frac{|\tilde{p}|}{2} +
u_0[|\tilde{p}|]\big{]}(\zeta_2 - \zeta_1)}\times
$$

$$
\sum_{l = 0}^{\infty}\sum_{m = -l}^{l}Y_{lm}(i\tilde{\theta}_2,
\phi_2)Y^*_{lm}(i\tilde{\theta}_1, \phi_1)
\int_{0}^{\infty}dw\frac{e^{i\frac{\pi}{\beta}w(\zeta_2 -
\zeta_1)}}{\pi^2(u_0[|\tilde{p}|] + w)^2 - 4(l^2 + l + 1)}
$$

$$
+ \frac{i}{\beta}
\frac{1}{2\pi}\frac{e^{i\frac{\pi}{\beta}u_0[0](\zeta_2 -
\zeta_1)}}{\sqrt{\zeta_1\zeta_2}}\int_{\frac{\pi}{2}}^{\frac{3\pi}{2}}d\theta
e^{i\theta}e^{i\frac{3\pi}{4\beta}e^{-i\theta}(\zeta_2 -
\zeta_1)}\times
$$

$$
\sum_{l = 0}^{\infty}\sum_{m = -l}^{l}Y_{lm}(i\tilde{\theta}_2,
\phi_2)Y^*_{lm}(i\tilde{\theta}_1, \phi_1)
\int_0^{\infty}dw\frac{e^{-\frac{\pi}{\beta}we^{-i\theta}(\zeta_2 -
\zeta_1)}}{\pi^2(u_0[0]e^{i\theta} + iw)^2 + 4(l^2 + l + 1)}
$$

$$
- \frac{i}{\beta}\frac{e^{\frac{3\pi}{4\beta}(\zeta_2 -
\zeta_1)}}{\sqrt{\zeta_1\zeta_2}}\int_{\epsilon \rightarrow
0}^{\infty}d|\tilde{p}|\frac{1}{1 - e^{-2\pi|\tilde{p}|}}
e^{i\frac{|\tilde{p}|}{\beta}(\xi_2 -
\xi_1)}e^{-i\frac{\pi}{\beta}\big{[}\frac{|\tilde{p}|}{2} +
u_0[|\tilde{p}|]\big{]}(\zeta_2 - \zeta_1)}\times
$$

$$
\sum_{l = 0}^{\infty}\sum_{m = -l}^{l}Y_{lm}(i\tilde{\theta}_2,
\phi_2)Y^*_{lm}(i\tilde{\theta}_1, \phi_1)
\int_{0}^{\infty}dw\frac{e^{-i\frac{\pi}{\beta}w(\zeta_2 -
\zeta_1)}}{\pi^2(u_0[|\tilde{p}|] + w)^2 - 4(l^2 + l + 1)}
$$

$$
- \frac{i}{\beta}\frac{e^{-\frac{3\pi}{4\beta}(\zeta_2 -
\zeta_1)}}{\sqrt{\zeta_1\zeta_2}}\int_{\epsilon \rightarrow
0}^{\infty}d|\tilde{p}|\frac{1}{1 - e^{-2\pi|\tilde{p}|}}
e^{i\frac{|\tilde{p}|}{\beta}(\xi_2 -
\xi_1)}e^{-i\frac{\pi}{\beta}\big{[}\frac{|\tilde{p}|}{2} +
u_0[|\tilde{p}|]\big{]}(\zeta_2 - \zeta_1)}\times
$$

$$
\sum_{l = 0}^{\infty}\sum_{m = -l}^{l}Y_{lm}(i\tilde{\theta}_2,
\phi_2)Y^*_{lm}(i\tilde{\theta}_1, \phi_1)
\int_{0}^{\infty}dw\frac{e^{i\frac{\pi}{\beta}w(\zeta_2 -
\zeta_1)}}{\pi^2(u_0[|\tilde{p}|] - w)^2 - 4(l^2 + l + 1)}
$$

$$
- \frac{i}{\beta}
\frac{1}{2\pi}\frac{e^{-i\frac{\pi}{\beta}u_0[0](\zeta_2 -
\zeta_1)}}{\sqrt{\zeta_1\zeta_2}}\int_{\frac{\pi}{2}}^{\frac{3\pi}{2}}d\theta
e^{i\theta}e^{-i\frac{3\pi}{4\beta}e^{-i\theta}(\zeta_2 -
\zeta_1)}\sum_{l = 0}^{\infty}\sum_{m =
-l}^{l}Y_{lm}(i\tilde{\theta}_2, \phi_2)Y^*_{lm}(i\tilde{\theta}_1,
\phi_1)\times
$$

\begin{equation}
\int_0^{\infty}dw\frac{e^{-\frac{\pi}{\beta}we^{-i\theta}(\zeta_2 -
\zeta_1)}}{\pi^2(u_0[0]e^{i\theta} - iw)^2 + 4(l^2 + l + 1)}
\end{equation}
with a range of validity for $ D^{(int)}(x_2 - x_1)$ specified  by

\begin{equation}
0 < |\rho_{int}|^2 \leq \frac{\beta^2}{100}
\end{equation}
and accordingly, in terms of $ r$, by

\begin{equation}
1.980M \leq r \leq 2M
\end{equation}

The derivation of Hawking radiation has been established as a
consequence of (7) and (11) in \cite{Tsoupros}. The essence of the
situation is depicted in Fig.1 and will be briefly outlined. In the
context of the pair-related approach described in the introduction
the negative-energy (anti)-particle advances toward the future
singularity forward in its proper time \footnote{Backward, in terms
of the Schwarzschild time-like coordinate $ r < 2M$.}. Equivalently,
a positive-energy particle emerges from the future singularity
and transits backward in its proper time across the event horizon. 
A quantum of the scalar field $ \Phi$ emerges at some point $ B$ of
the future singularity, transits as a positive-energy state backward
in its proper time across the future event horizon $ H^+$ where it
is scattered by the background curvature at some point $ C$ in the
vicinity of the latter and continues its transition forward in time
always as a positive-energy state before it registers its existence
at space-time point $ A \equiv (t'_1, r'_1, \theta'_1, \phi'_1)$ on
a detector whose world line is characterised by the fixed radial
coordinate $ r'_1$ \footnote{Actually scalar fields distinguish
neither between the two arrows of time nor, for that matter, between
positive and negative energy states. This fact, however, is
irrelevant to this heuristic picture.}. ``Forward in time" now has
the significance of forward in both, the particle's proper time and
the Schwarzschild coordinate time $ t$. As $ t$ is, at once, the
distant observer's proper time such a significance is of essence to
the boundary condition of forward-time propagation for positive
energy states which shall be analyzed in due course. The value $
r'_1$ falls within the range stated in (9). The syncopated segment $
BZ$ represents quantum propagation at radial values below the lower
bound $ r_1 = 1.980M$ in (13). The segment which is depicted by the
continuous line from $ Z \equiv (t_1, r_1, \theta_1, \phi_1)$ to
some point on the future event horizon represents the contribution
to the interior propagator which stems from the range of validity
stated in (13). Consequently, the segment $ ZCA$ represents the
massless conformal scalar propagator whose singular part is given by
(11) in the interior region of the Schwarzschild black hole and by
(7) altogether in the corresponding exterior region respectively.

\begin{figure}[h]
\centering\epsfig{figure=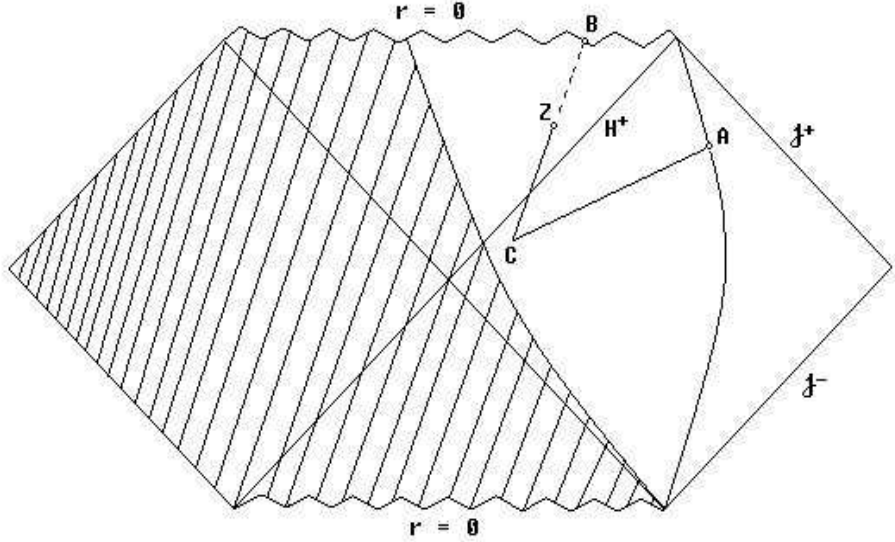, height =105mm,width=141mm}
\caption{Particle production by the Schwarzschild black hole in the
Hartle-Hawking vacuum state. The segment $ ZCA$ corresponds to
radial distances from the future event horizon $ H^+$ within which
the analytic Green functions in \cite{Tsoupros} and \cite{George}
coincide with the exact massless scalar propagator. The fixed radial
distance of the observer whose world line is also depicted in the
diagram lies within the range of that coincidence.}
\end{figure}

The probability for the emergence and transition of the quantum
particle from point $ Z$ in the interior region to the observation
point $ A$ is the square of the magnitude of the transition
amplitude from $ Z$ to $ A$. That transition amplitude is itself
equivalent to the functional integral over all paths which intercept
$ Z$ and $ A$  on condition that they do not pass through the
interior of the collapsing matter or extend to past infinity $ J^-$
\cite{Stephen Hawking}. As the Schwarzschild metric is regular on
the event horizon when expressed in Kruskal coordinates the
associated action functional is rigorously defined. The functional
integral can thus be given the concrete meaning which it has in (1)
through the analytic continuation of the Kruskal coordinates to a
domain in which the Schwarzschild metric has signature +4.

The transition amplitude between $ Z$ and $ A$ expressed in terms of
the scalar propagators in the exterior region and in the interior
region respectively is a superposition over a complete set of states
defined on the future event horizon $ H^+$ of the extended Kruskal
geometry \cite{Tsoupros}. Consequently, in the Euclidean sector of
the Schwarzschild metric which is respectively expressed by (6) and
by (10) in the context of (4) the transition amplitude $ D(x'_1 -
x_1)$ between the interior space-time point $ (i\xi_1, i\zeta_1,
i\theta_1, \phi_1)$ associated with $ Z$ and the exterior space-time
point $ (\tau'_1, \rho'_1, \theta'_1, \phi'_1)$ associated with $ A$
is necessarily expressed in terms of (7) and (11) through
superposition over a complete set of states defined on the
two-dimensional sphere at $ \rho = 0 \Leftrightarrow r = 2M$. That
superposition is itself applied on the understanding that - with the
angular dependence on $ \theta$ and $ \phi$ suppressed - the
coordinate singularity has caused through (4) the two sections which
respectively correspond to $ r \geq 2M$ and to $ r \leq 2M$ to be
connected only at the point $ \rho = 0$.

In this context $ D(x'_1 - x_1)$ reduces to \footnote{This corrects
an obviously inadvertent and irrelevant typesetting error in the
corresponding expression in \cite{Tsoupros}.}

$$
<\rho'_1, \theta'_1, \phi'_1, \tau'_1|i\zeta_1, i\theta_1, \phi_1,
i\xi_1> =
$$

$$
(2M)^2\int_{-1}^{1}dcos\theta_2\int_{0}^{2\pi}d\phi_2
\big{[}D_{as}(x'_1 - x_2)D_{as}^{(int)}(x_2 - x_1)\big{]} +
$$

\begin{equation}
(2M)^2\int_{-1}^{1}dcos\theta_2\int_{0}^{2\pi}d\phi_2
\big{[}D_b(x'_1 - x_2)D_b^{(int)}(x_2 - x_1)\big{]}
\end{equation}
and formally establishes through its asymptotic part $ D_{as}(x'_1 -
x_1)$ that within a finite advance of his proper time the observer
registers at point $ A$ radiation emitted by the Schwarzschild black
hole.

Crucial to the derivation of black-hole radiation is the behaviour
which the singular parts $ D_{as}^{(int)}(x_2 - x_1)$ and $
D_{as}(x'_1 - x_2)$ respectively have as a consequence of the
background geometry's causal structure. Specifically, at $ r_2
\rightarrow 2M^- \Leftrightarrow \zeta_2 \rightarrow 0$ the two
time-independent terms in (11), being the only two non-vanishing
terms, combine in the singular part of the propagator in (14) with $
lim_{\rho_2 \rightarrow 0}D_{as}(x'_1 - x_2)$ in such a manner as to
result in \cite{Tsoupros}

$$
D_{as}(x'_1 - x_1) = (2M)^2
\int_{-1}^{1}dcos\theta_2\int_0^{2\pi}d\phi_2\big{[}\frac{i}{8\pi
M}\frac{1}{\sqrt{\zeta_1}}
\sum_{l=0}^{\infty}\sum_{m=-l}^{l}Y_{lm}(i\tilde{\theta}_2,
\phi_2)Y_{lm}^{*}(i\tilde{\theta}_1, \phi_1)\times
$$

$$
[e^{-i\frac{\pi}{4M}u_0[0]
\zeta_1}\int_{\frac{\pi}{2}}^{\frac{3\pi}{2}}d\theta
e^{i\theta}e^{-i\frac{3\pi}{16M}e^{-i\theta}\zeta_1}
\int_0^{\infty}dw\frac{e^{\frac{\pi}{4M}we^{-i\theta}
\zeta_1}}{\pi^2(u_0[0]e^{i\theta} + iw)^2 + 4(l^2 +l + 1)} -
$$

$$
e^{i\frac{\pi}{4M}u_0[0]
\zeta_1}\int_{\frac{\pi}{2}}^{\frac{3\pi}{2}}d\theta
e^{i\theta}e^{i\frac{3\pi}{16M}e^{-i\theta}\zeta_1}
\int_0^{\infty}dw\frac{e^{\frac{\pi}{4M}we^{-i\theta}
\zeta_1}}{\pi^2(u_0[0]e^{i\theta} - iw)^2 + 4(l^2 +l + 1)}]\times
$$

\begin{equation}
\frac{1}{M^2}\frac{1}{32\pi^2}\frac{1}{\sqrt{\rho'_1}}\sum_{k =
0}^{\infty}\sum_{p
> p_0 > 0}^{\infty}e^{i\frac{p}{4M}n(8\pi
M)}e^{-i\frac{p}{4M}\tau'_1}\frac{2k + 1}{p}P_k(cos\gamma)\big{]}
\end{equation}
where $ cos\gamma = cos\theta_2cos\theta_1 +
sin\theta_2sin\theta_1cos(\phi_2 - \phi_1)$ and $ \tau_2 \rightarrow
\infty \Rightarrow n \rightarrow \infty$, a fact which will be
thoroughly analyzed in what follows. In \cite{Tsoupros} the singular
part of the exterior propagator as appears in the last line of (15)
also involves a multiplicative factor $ C << 1$. The origin of that
factor has been analyzed in \cite{George} and its presence in
\cite{Tsoupros} characterises the exterior propagator's behaviour
when one end-point of propagation is specified arbitrarily close to
the event horizon. In $ D_{as}(x'_1 - x_1)$ that factor cancels
essentially against the corresponding factor inherent in the
boundary condition that (11) diverge as $ \zeta_2^{-\frac{1}{2}}$ at
$ \zeta_2 \rightarrow 0$.

The transition amplitude $ D(x'_1 - x_1)$ in (14) is - with $ R = 0$
- equivalent to (1) with

\begin{equation}
S[\Phi, \nabla_{\alpha}\Phi] = \frac{1}{2}\int
d^4x[-g(x)]^{\frac{1}{2}}
g^{\mu\nu}(x)\partial_{\mu}\Phi(x)\partial_{\nu}\Phi(x)
\end{equation}
on condition that the Hartle-Hawking vacuum state of the field $
\Phi(x)$ has been excited only to the stated quantum particle. The
semi-classical limit is, in fact, inherent in the transition
amplitude stated in (15) since in the exterior vicinity of the event
horizon it is $ \tau_2 \rightarrow \infty$ with $ \tau_2 = n(8\pi M)
~~~~; ~~~~ n \rightarrow \infty$. In the present physical context
this is, indeed, the essence of the semi-classical situation $ \hbar
\rightarrow 0$ which, as stated in the introduction, characterises
the pair-production approach in the vicinity of the event horizon
with significance which decreases as $ \rho'_1$ increases. In
effect, in the exterior vicinity of the event horizon the entire
transition amplitude $ D(x'_1 - x_1)$ in (14) reduces to (2). In
addition, the fact that in the exterior vicinity of the event
horizon the imaginary-time coordinate $ \tau_2$ is an integer
multiple of $ 8\pi M$ is a necessary and sufficient condition for
the thermal character of Hawking radiation at $ T = (8\pi M)^{-1}$.
This shall also be amply demonstrated in what follows.

No less important an aspect of black-hole radiation than the stated
semi-classicality is the state of very high energy in which - as
stated in the introduction - each emitted particle is in the
exterior vicinity of the event horizon. In turn, this attribute has
an immediate implication for the elliptic operator from whose
eigenvalues and eigenstates the massless conformal scalar propagator
was constructed in the Euclidean sector of the Schwarzschild metric
\cite{George}. That is, the dominant contribution to the
semi-classical limit of $ D(x'_1 - x_1)$ which describes the
amplitude for the emitted particle to be detected in a local frame
close to the event horizon comes from eigenvalues and eigenstates of
very high order. Consequently, in addition to reducing to (2) in
that frame, the entire transition amplitude $ D(x'_1 - x_1)$ for a
quantum of a scalar field also reduces to $ D_{as}(x'_1 - x_1)$.

In order to explore the behaviour of (15) and, consequently, of the transition
amplitude in (14) at the  stated semi-classical limit it is necessary to convert
the infinite series over $ p$ to a contour integral. In anticipation of that it
is

$$
\sum_{k = 0}^{\infty}\sum_{p > p_0}^{\infty}e^{i\frac{p}{4M}n(8\pi
M)}e^{-i\frac{p}{4M}\tau'_1}\frac{2k + 1}{p}P_k(cos\gamma) =
$$

\begin{equation}
e^{i\frac{p_0 + 1}{4M}n(8\pi M)}\sum_{k = 0}^{\infty}(2k +
1)P_k(cos\gamma)\sum_{p > p_0}^{\infty}\frac{1}{p}e^{i\frac{p - (p_0 +1)}{4M}n(8\pi
M)}e^{-i\frac{p}{4M}\tau'_1}
\end{equation}

Upon extending $ p \in R$ to $ \tilde{p} \in C$ the residue theorem implies
that (17) can be expressed as

\begin{equation}
\frac{1}{2i}e^{i\frac{p_0 + 1}{4M}n(8\pi M)}\sum_{k = 0}^{\infty}(2k +
1)P_k(cos\gamma)\oint_{c_{\tilde{p}}}d\tilde{p}\frac{(-1)^{\tilde{p}}}{sin{(\pi\tilde{p})}}
\frac{1}{\tilde{p}}e^{i\frac{\tilde{p} - (p_0 + 1)}{4M}n(8\pi
M)}e^{-i\frac{\tilde{p}}{4M}\tau'_1}
\end{equation}
where the infinite contour $ c_{\tilde{p}}$ encompasses all integers bigger than some
$ p_0 \geq 1$ as in Fig.2.

\begin{figure}[h]
\centering\epsfig{figure=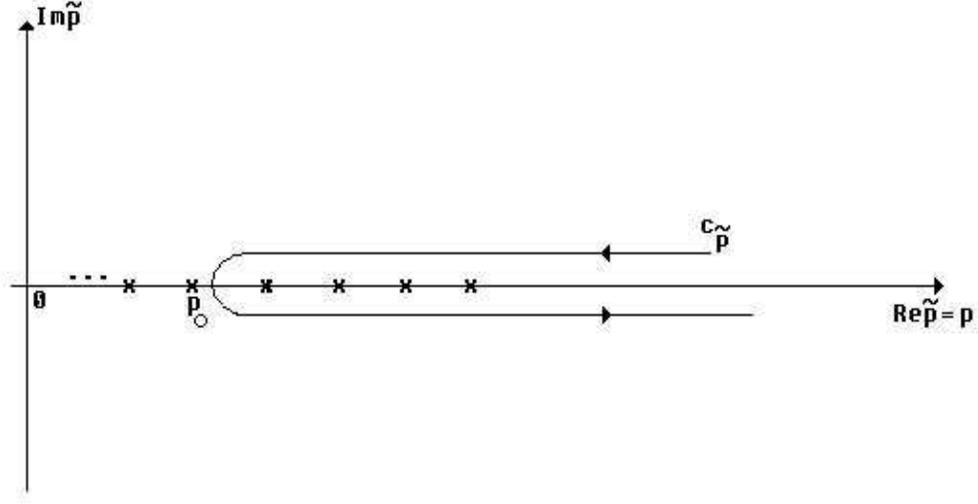, height =85mm,width=141mm}
\caption{The infinite contour $ c_{\tilde{p}}$}
\end{figure}

Inspection of the expression in (18) reveals that at $ n \rightarrow \infty$ the integrand oscillates
with infinite frequency unless $ \tilde{p} - (p_0 + 1) = |\tilde{p} - (p_0 + 1)|e^{i\theta}$ is, for all intents and purposes, a constant.
Consequently, the only contribution to (18) stems from the relevant segment of
$ c_{\tilde{p}}$'s semi-circle along which $ cos\theta \sim -1$.

Since

\begin{equation}
e^{i\frac{\tilde{p} - (p_0 + 1)}{4M}n(8\pi
M)} = e^{i\frac{|\tilde{p} - (p_0 + 1)|}{4M}n(8\pi M)cos\theta}e^{- \frac{|\tilde{p} - (p_0 + 1)|}{4M}n(8\pi M)sin\theta} 
\end{equation}
it follows that at $ n \rightarrow \infty$ (17) reduces to

$$
\frac{1}{2i}e^{i\frac{p_0 + 1}{4M}n(8\pi M)}\sum_{k = 0}^{\infty}(2k
+ 1)P_k(cos\gamma)\times
$$

\begin{equation}
i|\tilde{p} - (p_0 + 1)|\int_{\pi - \epsilon}^{\pi +
\epsilon}d\theta e^{i\theta}
\frac{(-1)^{\tilde{p}}}{sin{(\pi\tilde{p})}} \frac{1}{\tilde{p}}
e^{i\frac{|\tilde{p} - (p_0 + 1)|}{4M}n(8\pi M)cos\theta}e^{-
\omega'(8\pi M)}e^{-i\frac{\tilde{p}}{4M}\tau'_1}
\end{equation}
with the value of the positive constant $ \epsilon$ being close to zero and with

\begin{equation}
\omega' = \frac{|\tilde{p} - (p_0 + 1)|}{4M}(nsin\theta)
\end{equation}

For $ \pi - \epsilon < \theta < \pi + \epsilon$ it is $ sin\theta
\sim \pi - \theta = \chi \in [-\epsilon, \epsilon]$. As $ \chi
\rightarrow 0^+$ the limit $ n \rightarrow \infty$ coincides with $
lim_{\chi \rightarrow 0^+} \frac{c}{\chi} ; c > 0$ and imposes the
restriction $ 0 < \omega' < \infty$. This restriction remains
identically valid as integration over $ \theta$ extends to negative
values of $ \chi$ which lie in a neighborhood of zero. This is the
immediate consequence of the positive (counterclockwise) orientation
of $ c_{\tilde{p}}$ and will be further analyzed below.

The replacement of (20) in (15) reduces the latter to

\begin{equation}
D_{as}(x'_1 - x_1) \sim e^{- 8\pi M\omega'}
\end{equation}

Since, as stated, in the local frame defined close to the event horizon $ D(x'_1 - x_1)$ reduces to
$ D_{as}(x'_1 - x_1)$ (22) also implies that in that frame it is

\begin{equation}
D(x'_1 - x_1) \sim e^{- 8\pi M\omega'}
\end{equation}
which is manifestly consistent with (2).

This allows for the evaluation of the probability $ \Gamma(x'_1 - x_1)$ that an
excitation of the massless conformal scalar field transit between $ x_1$ and $ x'_1$. Since the latter is
the product between $ D(x'_1 - x_1)$ and its conjugate expression it follows that

\begin{equation}
\Gamma(x'_1 - x_1) \sim e^{- 8\pi M\omega} ~~~~ ; ~~~~ \omega =
2\omega'
\end{equation}

The positive quantity $ \omega$ in (24) can now be identified with
the energy of the emitted particle. Such an identification is the
necessary consequence of the positive orientation which was imposed
on the contour $ c_{\tilde{p}}$ in (18). A negative orientation of $
c_{\tilde{p}}$ would have resulted in a physically unacceptable
negative $ \omega$. It can be seen, for that matter, that the
positive orientation of $ c_{\tilde{p}}$ signifies a boundary
condition which stems from the physical demand that the emitted
particles correspond to positive-energy states which propagate
forward in time. The formal statement made in connection to (15) to the
effect that the Schwarzschild black hole radiates must be
supplemented with the stated boundary condition. It is that boundary
condition which reflects a non-vanishing contribution to contour
integration at $ \tau_2 \rightarrow \infty$ while enforcing the stated
demand of forward-time propagation for positive-energy states. In turn,
that boundary condition which is essentially imposed on the future event
horizon $ H^+$ is additional to the boundary condition for vanishing
propagation at spatial infinity imposed on the solution to the Green
equation in \cite{George} and completes the construction of the
massless conformal scalar propagator which was achieved in
\cite{George} and \cite{Tsoupros} in the exterior and interior
region of the Schwarzschild black hole respectively.

It should be remarked at this point that - in line with expectations
- the identification of $ \omega$ with the energy of the emitted
particle advances (23) into coincidence with (2) since $ 8\pi
M\omega'$ now has dimensions of action and - up to the logarithm of
the entire factor which multiplies the exponential - it is, indeed,
the Euclidean semi-classical action for the particle emitted by the
Schwarzschild black hole in the Hartle-Hawking state.

An issue which merits attention in its own right is that raised in
\cite{Predictability}. S. Hawking has effectively argued that it is
because the particles which constitute the black-hole radiation
emanate from the interior of the black hole about which an external
observer has no knowledge that such an observer can not predict the
amplitudes for them to be emitted. He can only predict the
probabilities of emission without the phases. At first sight it
would appear that the result expressed by (15) and (23) contradicts
that statement. Such an appearance is deceptive for two reasons. The
first is inherent in the factor $ u_0[0]$, explicitly featured in
(15). The origin of that factor can be traced to the separation of
the exterior propagator in a singular part and a boundary part and
has been thoroughly analyzed in \cite{George}. As that separation is
arbitrary so is $ u_0[0]$. Any change in the latter necessarily
effects a change in the boundary part of the exterior propagator in
such a manner as to leave the exterior propagator intact. Not so,
however, in the case of particle emission from the Schwarzschild
black hole. As stated, the high-energy state of the strongly
semi-classical particles emitted in the exterior vicinity of the
event horizon brings the entire transition amplitude for each
quantum of a scalar field into coincidence with that amplitude's
singular part. Consequently, $ u_0[0]$ can no longer be arbitrary.
The determination of $ u_0[0]$ can now only be the result of a
boundary condition imposed on the singular part of the exterior
propagator. In turn, that boundary condition necessarily relates to
the physical context in the interior region about which the external
observer has no knowledge and is, for that matter, unattainable.

The second ambiguity inherent in the transition amplitude expressed
by (15) and (23) relates to the fact that in the immediate exterior
vicinity of the event horizon the imaginary-time coordinate $
\tau_2$ is an integer multiple of $ 8\pi M$. The precise
significance of that statement is $ \tau_2 = (n + c)(8\pi M)
~~~~;~~~~ c \in R$ with $ lim_{\rho_2 \rightarrow 0}n = \infty$. As
the preceding analysis reveals the presence of the constant $ c$
will introduce yet another multiplicative phase factor in the entire
factor which multiplies the exponential in (23). That factor is
indeterminate since knowledge of the value of $ c$ can, again, only
be elicited by the physical context in the interior region and is,
for that matter, unattainable by the exterior observer.

In conclusion, the transition probability in (24) signifies a Boltzmann
factor for a particle of energy $ \omega$ at Hawking temperature
$ T = \frac{1}{8\pi M}$. That, in turn, expresses the thermal character
of Hawking radiation. This result has been established in a local frame
close to the Schwarzschild event horizon. The Schwarzschild radial
coordinate $ r$ associated with that frame falls within the range of
validity stated in (9). Through further superposition over a complete
set of states defined on a hypersurface which includes the event of that
particle's registration in that frame the same result can trivially be
extended to the ``inertial" observer at $ r \rightarrow \infty$.

{\bf III. The Detraction}

The preceding analysis can now be extended to the detraction from
the thermal character in which the tunneling approach in
\cite{Wilczek} and \cite{Parikh} results. As stated in the
introduction that approach has the distinctive feature of treating
each emitted quantum particle as the result of such tunneling across
the event horizon as is characterised by a barrier which crucially
depends on the tunneling particle itself. Specifically, whereas in
the original approach analyzed in the previous section the emission
of a particle of energy $ \omega$ necessarily corresponds to only
one entangled quantum state (that is, to only one pair of quantum
particles) which originates just above the event horizon in a fixed
exterior geometry, as in Fig.1, the approach in \cite{Wilczek} and
\cite{Parikh} involves two possible entangled quantum states one of
which corresponds to the description in Fig.1 with emitted energy
equal to $ \omega$ and the other one of which originates just below
the event horizon and corresponds to the tunneling of a particle of
energy $ \omega$ forward in the particle's proper time. In the
context of either entangled state energy conservation causes the
event horizon to contract to a smaller Schwarzschild radius.

The event horizon in its initial location and in the final location
to which it contracts yields respectively two null hypersurfaces
each of which qualifies for a classical turning point. The
separation of these two hypersurfaces corresponds to the classically
inaccessible region through which the semi-classical particle must
tunnel as a condition for its emission. 
For that matter, this approach to black-hole
radiation actively involves the event horizon and, indeed, causes
its exterior vicinity to be stationary but not static. This renders
the expression of the Schwarzschild metric in Schwarzschild
coordinates in (3) inadequate for the description of such a
situation. A particularly suitable coordinate system which renders
that stationary character of the exterior region manifest and which,
at once, is regular at $ r = 2M$ is that of Gullstrand-Painlev\'{e}
coordinates

\begin{equation}
ds^2 = -(1 - \frac{2M}{r})dt^2 + 2\sqrt{\frac{2M}{r}}dtdr + dr^2 +
r^2d\Omega^2
\end{equation}
with the temporal coordinate $ t$ being distinct from the
corresponding Schwarzschild radial coordinate in (3).

The self-gravitating character of the emitted particle of energy $
\omega$ eventually modifies the background geometry to
\cite{Wilczek}

\begin{equation}
ds^2 = -\big{(}1 - \frac{2(M - \omega)}{r}\big{)}dt^2 +
2\sqrt{\frac{2(M - \omega)}{r}}dtdr + dr^2 + r^2d\Omega^2
\end{equation}

An essential digression is in order at this point. In Planck units $
M$ has dimensions of length. For that matter, so does $ \omega$ in
(26). Consequently, $ \omega$ in (26) does not coincide with $
\omega$ in (24) since - as (21) explicitly shows - the latter has
dimensions of mass. In spite of that, the $ \omega$ which appears in
(26) unduly multiplies the expression $ M - \frac{\omega}{2}$ in the
final result for the emission probability $ \Gamma \sim e^{-8\pi
\omega(M - \frac{\omega}{2})}$ and in all associated expressions
which appear in \cite{Wilczek} resulting in length-dimensionality of
order two in the exponent. Taken at ``face value" such a result is
incorrect as it identically contradicts the fact that the exponent
which it features must have dimensions of action. A careful
consideration of the mass dimensionality in the derivation cited in
\cite{Wilczek} reveals, indeed, that the $ \omega$ which multiplies
$ M - \frac{\omega}{2}$ in that exponent is of mass-dimensionality
of order one and coincides, for that matter, with the $ \omega$ in
(24) whereas the $ \omega$ which appears in $ M - \frac{\omega}{2}$
and in (26) is of length dimensionality of order one. In order to
bring my result into line with that in \cite{Wilczek} I shall, in
what follows, adhere to the stated convention which the authors of
\cite{Wilczek} have made.

The contraction of the event horizon is now manifest in (26). A
particle of positive energy $ \omega$ emerges at the future
singularity and transits backward in its proper time to a point at $
r_{in} = 2M - \epsilon ; \epsilon \rightarrow 0^+$ whence it tunnels
through the contracting future event horizon forward in its proper
time to $ r_{out} = 2(M - \omega) + \epsilon ; \epsilon \rightarrow
0^+$. The separation between the event horizon at Schwarzschild
radius equal to $ 2M$ and that at $ 2(M - \omega)$ defines the
classically inaccessible region of this tunneling process. The other
entangled state associated with the emission of a particle of energy
$ \omega$ emerges at space-time point $ C$ just above the future
event horizon and involves two semi-classical particles each of
which is of positive energy $ \omega$. Upon crossing the event
horizon inward one of the two particles reverses the sign of its
energy to $ - \omega$ with respect to spatial infinity.
Equivalently, that particle emerges at the future singularity in a
positive-energy state and propagates backward in its proper time
until space-time point $ C$ where it is scattered by the background
curvature in forward-time propagation to the observer at space-time
point $ A$. Again, energy conservation causes the event horizon to
contract from its initial location at Schwarzschild radius equal to
$ 2M$ to a final location at Schwarzschild radius equal to $ 2(M -
\omega)$. The distinction between these two entangled states has no
operational significance for the observer whose world line is
depicted in Fig.1. Since both entangled quantum states contribute to
the total probability of emission his observation at space-time
point $ A$ consists in the particle of energy $ \omega$ itself and
in the above-stated contraction of the event horizon.

The evaluation of the probability that a scalar particle of energy $
\omega$ be emitted in a physical context which is characterised by
an evolving event horizon - equivalently, the evaluation of that
particle's transition probability - in terms of the propagators in
\cite{George} and \cite{Tsoupros} can be accomplished through a
careful analysis of the relation between that evolution and those
propagators. To that end, all aspects of the analytical procedure
which was, in the previous section, applied to (15) will be
re-assessed in the present physical context. In what follows the
emitted scalar quantum will be treated as a spherical shell. Such an
approach is justified by the fact that the emission of a
scalar-field configuration by the Schwarzschild black hole occurs
primarily as an s-wave \cite{KrausWilczek}.

As observed from spatial infinity the sum-total of the hole's mass
and that corresponding to the energy of the emitted spherical shell
is constant while the mass of the hole varies according to the
energy of the emitted shell\footnote{In the Hamiltonian approach the
constraints in the theory allow for time-dependence in the mass of
the spherical shell and in that of the black hole
\cite{KrausWilczek}.}. By extension, such will also be the case in
the frame of the observer at $ A$ in Fig.1. Energy conservation is
the underlying cause of that situation as a consequence of which in
the course of emission the background space-time geometry becomes
dynamic in the exterior vicinity of the event horizon and the
massless semi-classical particle transits on a null geodesic. That
geodesic corresponds to the metric in (26) and describes the motion
of a particle of energy $ \omega$ in the stationary region of
space-time bounded by the original location of the contracting event
horizon at $ r_s = 2M$ and by its final location at $ r'_s = 2(M -
\omega)$ \cite{Wilczek}. This is the essence of the situation in the
context of either entangled state.

In the transition amplitude which corresponds to (15) the quantity $
M$ determines the background curvature of the Schwarzschild
black-hole geometry on which the quantum scalar excitation
propagates. In the present physical context the background curvature
is determined by the constant total mass $ (M - \omega) + \omega$ as
observed at $ A$. Since this coincides with the mass of the
Schwarzschild black hole prior to the emission of the spherical
shell so does the quantity $ M$ in the transition amplitude which
corresponds to (15). In fact, all space-time points in the exterior
region retain their $ r$-values in the course of the shell's
propagation. That concerns, all the same, the set of points at $ r =
2M$ which defined the event horizon prior to the emission. As a
consequence, in the event of emission $ \rho'_1$ also retains the
value specified by (4).

In the present context, for that matter, the mathematical expression
in (15) remains - at this stage - identically the same. If the
analytical evaluation in the previous section were to be applied in
this context the same arbitrary positive constant $ \omega'$ would
emerge. This, directly calls the relation of $ \omega'$ to the
emitted energy $ \omega$ into question. The present physical
context, which is centrally determined by energy conservation,
involves two entangled quantum states as opposed to the one
represented in Fig.1. Both of these states contribute to the
transition amplitude. This fact, however, concerns exclusively the
difference in the probability for emission of energy equal to $
\omega$ in the two distinct contexts. Since in either context that
probability concerns a particle of energy equal to $ \omega$ it
follows that, pursuant to (24), the transition amplitude which is
determined by the present physical context is also characterised by
$ \omega' = \frac{\omega}{2}$.

The expression in (24) signifies the characteristic Boltzmann factor
for the thermal emission of a particle of energy $ \omega$ at
temperature $ T = [8\pi M]^{-1}$ and is, in the present context,
physically inconsistent in view of the emitted particle's
self-gravitating character. Such an inconsistency is a direct
consequence of the fact that the interpretation of the physical
content of (15) is incomplete. In order to properly take the
reduction of the hole's mass by $ \omega$ into consideration use
must be made of the fact that, prior to the emission, in the
vicinity of the event horizon it is $ \tau_2 = n(8\pi M)~~~~~ ; ~~~~
n \rightarrow \infty$. The replacement of the temporal period which
appears in this mathematical statement by any positive constant
leaves the latter invariant. In the present physical context which
is centrally characterised by a contracting event horizon that
positive constant will necessarily be determined by the constraint
which energy conservation imposes on the system. The latter demands
that the initial mass $ M$ of the black hole be reduced by $ \omega$
in the event of emission of energy equal to $ \omega$. Since - as
shown above - the emitted energy $ \omega$ corresponds to $
\frac{\omega}{2}$ in the transition amplitude it follows that the
hole's mass $ M$ must, in the temporal period $ 8\pi M$ which
appears in (15), be reduced by the same amount. The replacement of $
8\pi M$ by $ 8\pi(M - \frac{\omega}{2})$ does not contradict the
stated constancy of the total mass $ M$ because the temporal period
in the Euclidean sector of the Schwarzschild metric is determined
exclusively by the mass of the black hole, which - as stated - is
variable.

Equivalently, the demand for energy conservation must naturally be
imposed on the Euclidean semi-classical action which - pursuant to
the analysis in section II - is equal to $ \frac{\omega}{2}(8\pi M)$
and inherent in (15). Since, in that action, the emitted energy $
\omega$ corresponds to $ \frac{\omega}{2}$ the expression $ n(8\pi
M)$ which appears in the temporal sector of (15) must, as a result
of the demand for conservation of energy, be replaced by $ n8\pi (M
- \frac{\omega}{2})$.

In effect, the constraint imposed on the transition amplitude
between points $ Z$ and $ A$ by the demand for conservation of
energy yields

$$
D(x'_1 - x_1) =
\int_{-1}^{1}dcos\theta_2\int_0^{2\pi}d\phi_2\big{[}\frac{i}{8\pi
M}\frac{1}{\sqrt{\zeta_1}}
\sum_{l=0}^{\infty}\sum_{m=-l}^{l}Y_{lm}(i\tilde{\theta}_2,
\phi_2)Y_{lm}^{*}(i\tilde{\theta}_1, \phi_1)\times
$$

$$
[e^{-i\frac{\pi}{4M}u_0[0]
\zeta_1}\int_{\frac{\pi}{2}}^{\frac{3\pi}{2}}d\theta
e^{i\theta}e^{-i\frac{3\pi}{16M}e^{-i\theta}\zeta_1}
\int_0^{\infty}dw\frac{e^{\frac{\pi}{4M}we^{-i\theta}
\zeta_1}}{\pi^2(u_0[0]e^{i\theta} + iw)^2 + 4(l^2 +l + 1)} -
$$

$$
e^{i\frac{\pi}{4M}u_0[0]
\zeta_1}\int_{\frac{\pi}{2}}^{\frac{3\pi}{2}}d\theta
e^{i\theta}e^{i\frac{3\pi}{16M}e^{-i\theta}\zeta_1}
\int_0^{\infty}dw\frac{e^{\frac{\pi}{4M}we^{-i\theta}
\zeta_1}}{\pi^2(u_0[0]e^{i\theta} - iw)^2 + 4(l^2 +l + 1)}]\times
$$

\begin{equation}
\frac{1}{8\pi^2}\frac{1}{\sqrt{\rho'_1}}\sum_{k = 0}^{\infty}\sum_{p
> p_0 > 0}^{\infty}e^{i\frac{p}{4M}n8\pi
(M - \frac{\omega}{2})}e^{-i\frac{p}{4M}\tau'_1}\frac{2k +
1}{p}P_k(cos\gamma)\big{]}
\end{equation}

It can be seen, for that matter, that the reduction of the hole's
mass effectively defines a new temporal period which reflects the
associated contraction of the event horizon and, thereby, the
dynamic character of the background geometry as a consequence of
vacuum activity. Clearly, the semi-classical propagation along the
null geodesics between $ r_{in} = 2M - \epsilon$ and $ r_{out} = 2(M
- \omega) + \epsilon$  and between $ r_{out} = 2M + \epsilon$ and $
r_{in} = 2(M - \omega) - \epsilon$ respectively for either entangled
quantum state is inherent in (27) and translates to the reduction of
the temporal period from $ 8\pi M$ to $ 8\pi (M- \frac{\omega}{2})$.

The transition amplitude in (27) is now consistent with energy
conservation, with the concomitant consequence of a contracting
event horizon and with a Schwarzschild geometry which corresponds to
a constant total mass in the course of emission. Consequently, in
the context of the demand for energy conservation (27) is the
correct expression for the massless conformal scalar propagator
between points $ Z$ and $ A$.

The formal procedure which was applied to (15) may now be,
consistently, applied to (27). That procedure reduces the transition
amplitude in (27) to

\begin{equation}
D(x'_1 - x_1) \sim e^{-(\frac{\omega}{2})8\pi (M -
\frac{\omega}{2})}
\end{equation}

The expression in (27) - and accordingly that in (28) - is the
desired transition amplitude for the emission of a self-gravitating
particle of energy $ \omega$ by the Schwarzschild black hole of mass
$ M$. In addition to being consistent with energy conservation the
expression in (28) also naturally reduces to that in (23) in the
event that $ \omega << M$. Crucial to the derivation of the
transition amplitude in (27) has been the demand for transition
between states of the same total energy.

The transition probability $ \Gamma(x'_1 - x_1)$ for an excitation
of the massless conformal scalar field between $ x_1$ and $ x'_1$ is
the product between $ D(x'_1 - x_1)$ and its conjugate expression.
From (28) it follows that

\begin{equation}
\Gamma(x'_1 - x_1) \sim e^{-8\pi \omega(M - \frac{\omega}{2})}
\end{equation}
which coincides with the result in \cite{Wilczek}.

The detraction from the thermal character of Hawking radiation
enforced by energy conservation is, in (29), manifest in the $
\omega^2$-term. As was also the case in the thermal context of the
previous section crucial to the derivation of this result has been
the boundary condition of positive orientation for the contour along
which the corresponding integration occurs in the complex plane. It
is invariably that boundary condition which enforces the physical
demand that the emitted particles correspond to positive-energy
states which propagate forward in time.

{\bf IV. Conclusions}

The propagator for a conformal scalar field in the Hartle-Hawking
state was developed in \cite{George} as an explicitly analytic
function of the background geometry for a finite range of values of
the Schwarzschild radial coordinate above the Schwarzschild radius.
That propagator was analytically extended into the interior region
of the Schwarzschild black hole in \cite{Tsoupros} in which it was
also established that the boundary condition which the causal
structure of the Schwarzschild black-hole space-time imposes on the
propagator in the interior region results in particle production by
the Schwarzschild black hole. The entire analysis in this project
has established that the particle production which the massless
conformal scalar propagator in \cite{George} and \cite{Tsoupros}
describes corresponds to Hawking radiation of thermal character if
the exterior background geometry is treated as a fixed background to
quantum propagation and to Hawking radiation which detracts from
thermality if energy conservation is enforced. Central to the
derivation of Hawking radiation and its character in either case is
the semi-classical limit $ \hbar \rightarrow 0$ which characterises
each emitted scalar quantum in the exterior vicinity of the event
horizon. That, in turn, imposes the condition that in the exterior
vicinity of the event horizon the imaginary temporal coordinate be,
in the thermal case, an integer multiple of its period $ 8\pi M$.
Although this statement remains formally valid if the radiation
detracts from thermality the demand for energy conservation enforces
the condition that in the vicinity of the event horizon the
imaginary temporal coordinate be, instead, an integer multiple of
the effective period $ 8\pi (M - \frac{\omega}{2})$ for each emitted
quantum of energy $ \omega$.

Crucial to the derivation of Hawking radiation from the scalar
propagators in \cite{George} and \cite{Tsoupros} is the boundary
condition that the contour of integration in, what is essentially,
the energy complex plane be of positive orientation. That boundary
condition is imposed on the propagator in the exterior vicinity of
the event horizon and enforces the physical demand that particles
emitted by the black hole correspond to positive-energy states which
propagate forward in time. That boundary condition completes, for
that matter, the construction of the massless conformal scalar
propagator on the Schwarzschild black-hole space-time which was
initiated in \cite{George}. The above-stated boundary condition
imposed by space-time's causal structure in the interior region,
that imposed in the exterior vicinity of the event horizon by the
demand for forward-time propagation of positive-energy states and
the boundary condition of vanishing propagation at spatial infinity
are the three boundary conditions which characterise the scalar
propagator on the Schwarzschild black-hole geometry.

The derivation of Hawking radiation from the scalar propagators in
\cite{George} and \cite{Tsoupros} supports the case made by S. Hawking to
the effect that an external observer can only predict the probability for
particle emission but not the amplitude for that emission.

The scalar propagators in \cite{George} and, especially, in
\cite{Tsoupros} are particularly involved. However, the derivation
of Hawking radiation - both, of thermal character and of a character
which detracts from thermality - is indicative of the potential for
enormous simplification which the causal structure of space-time has
in any calculational context which describes conformal scalar
propagation on the Schwarzschild black-hole geometry.

\end{document}